\documentclass[twocolumn,english,prl,superscriptaddress]{revtex4}
\usepackage[T1]{fontenc}
\usepackage[latin9]{inputenc}
\usepackage{textcomp}
\usepackage{amstext}
\usepackage{graphicx}
\usepackage{amssymb}
\usepackage{esint}

\makeatletter
\@ifundefined{textcolor}{}
{%
 \definecolor{BLACK}{gray}{0}
 \definecolor{WHITE}{gray}{1}
 \definecolor{RED}{rgb}{1,0,0}
 \definecolor{GREEN}{rgb}{0,1,0}
 \definecolor{BLUE}{rgb}{0,0,1}
 \definecolor{CYAN}{cmyk}{1,0,0,0}
 \definecolor{MAGENTA}{cmyk}{0,1,0,0}
 \definecolor{YELLOW}{cmyk}{0,0,1,0}
 }

\makeatother

\usepackage{babel}

\begin{document}

\title{An accurate measurement of electron beam induced displacement cross
sections for single-layer graphene}

\author{Jannik C. Meyer}

\affiliation{Central Facility for Electron Microscopy, Group of Electron Microscopy
of Materials Science, University of Ulm, Albert Einstein Allee 11,
89081 Ulm, Germany}

\affiliation{University of Vienna, Department of Physics, Boltzmanngasse 5, 1090
Vienna, Austria}

\author{Franz Eder}

\affiliation{University of Vienna, Department of Physics, Boltzmanngasse 5, 1090
Vienna, Austria}

\author{Simon Kurasch}

\affiliation{Central Facility for Electron Microscopy, Group of Electron Microscopy
of Materials Science, University of Ulm, Albert Einstein Allee 11,
89081 Ulm, Germany}

\author{Viera Skakalova}

\affiliation{Max Planck Institute for solid state research, Heisenbergstr. 1,
70569 Stuttgart, Germany}

\affiliation{University of Vienna, Department of Physics, Boltzmanngasse 5, 1090
Vienna, Austria}

\author{Jani Kotakoski}

\affiliation{Department of Physics, University of Helsinki, P.O. Box 43, 00014
Helsinki, Finland}

\affiliation{University of Vienna, Department of Physics, Boltzmanngasse 5, 1090
Vienna, Austria}

\author{Hye Jin Park}

\affiliation{Max Planck Institute for solid state research, Heisenbergstr. 1,
70569 Stuttgart, Germany}

\author{Siegmar Roth}

\affiliation{Max Planck Institute for solid state research, Heisenbergstr. 1,
70569 Stuttgart, Germany}

\affiliation{WCU Flexible Electronics, School of Electrical Engineering, Korea
University, Seoul, Korea}

\author{Andrey Chuvilin}

\affiliation{Central Facility for Electron Microscopy, Group of Electron Microscopy
of Materials Science, University of Ulm, Albert Einstein Allee 11,
89081 Ulm, Germany}

\affiliation{CIC nanoGUNE Consolider, Av. de Tolosa 76, 20018, San Sebastian,
Spain and Ikerbasque, Basque Foundation for Science, 48011, Bilbao, Spain}

\author{Sören Eyhusen}

\affiliation{Carl Zeiss NTS GmbH, Carl-Zeiss-Strasse 56, 73447 Oberkochen, Germany}

\author{Gerd Benner}

\affiliation{Carl Zeiss NTS GmbH, Carl-Zeiss-Strasse 56, 73447 Oberkochen, Germany}

\author{Arkady V. Krasheninnikov}

\affiliation{Department of Physics, University of Helsinki, P.O. Box 43, 00014
Helsinki, Finland}

\affiliation{Department of Applied Physics, Aalto University, P.O. Box 1100, 00076
Aalto, Finland}

\author{Ute Kaiser}

\affiliation{Central Facility for Electron Microscopy, Group of Electron Microscopy
of Materials Science, University of Ulm, Albert Einstein Allee 11,
89081 Ulm, Germany}
\begin{abstract}
We present an accurate measurement and a quantitative analysis of
electron-beam induced displacements of carbon atoms in single-layer
graphene. We directly measure the atomic displacement ({}``knock-on'')
cross section by counting the lost atoms as a function of the electron
beam energy and applied dose. Further, we separate knock-on damage
(originating from the collision of the beam electrons with the nucleus
of the target atom) from other radiation damage mechanisms (e.g. ionization
damage or chemical etching) by the comparison of ordinary ($^{12}\textrm{C}$)
and heavy ($^{13}\textrm{C}$) graphene. Our analysis shows that a
static lattice approximation is not sufficient to describe knock-on
damage in this material, while a very good agreement between calculated
and experimental cross sections is obtained if lattice vibrations
are taken into account.
\end{abstract}
\maketitle
Radiation damage is one of the key limitations of high-resolution
transmission electron microscopy (HR-TEM) \cite{Egerton2004}. In
particular, the continuous improvements in instrumental resolution
\cite{Haider1998,Batson2002} inevitably entail increased doses per
area that need to be applied to a sample. The need for high doses
is further increased for new techniques such as single-atom or single-atomic-column
spectroscopy \cite{Batson1993,Kaiser2002,Muller2008,Suenaga2009},
atomic resolution electron tomography \cite{BarSadan2008}, or the
analysis of charge distributions from very high signal-to-noise ratio
HR-TEM images \cite{Meyer2011}. For light element materials, such
as carbon nanotubes \cite{Iijima1991,Iijima1993}, fullerenes \cite{Kroto1985},
graphene \cite{Novoselov2004,Geim2007}, boron nitride \cite{Pacile2008,Novoselov2005},
and probably many more, the dose limitation is particularly severe
for three reasons: First, it is obvious, that knock-on damage cross
sections will be higher for low atomic number elements \cite{Banhart1999}.
Second, the light elements produce less contrast than heavier elements,
so that even higher doses are needed to obtain a sufficient signal
to noise ratio. And third, most of the novel materials from light
elements, such as graphene or carbon nanotubes, appear in the form
of low dimensional allotropes that have only one or a few atoms in
a typical projection of a high-resolution image. While almost all
atomic spacings can in principle be resolved by the currently available
instrumentation, the question remains whether a sample is stable under
the beam until an image has been acquired. In spite of a wide range
of previous studies concerning irradiation damage in carbon nanostructures
\cite{LUCAS1964,Crespi1996a,Banhart1999,Smith2001,Krasheninnikov2005,Zobelli2007a,Molhave2007,Warner2009a},
a quantitative experimental determination of atomic displacement cross
sections for this important class of materials is absent. In fact,
only very few quantitative measurements of electron-beam induced displacement
cross sections \cite{CHERNS1976,Egerton2010} (beyond damage threshold
measurements \cite{Egerton1977,Zag1983}) can be found in the literature.
The understanding of irradiation effects is also important for targeted
irradiation-induced modifications of a material: For the case of graphene,
for example, a controlled introduction of vacancies and non-hexagonal
rings may lead to derived $sp^{2}$ hybridized carbon sheets with
specific properties \cite{Crespi1996,Terrones2000,Lusk2008,Fischbein2008,Kotakoski2011,Krasheninnikov2007}.

\begin{figure}
\includegraphics[width=1\columnwidth]{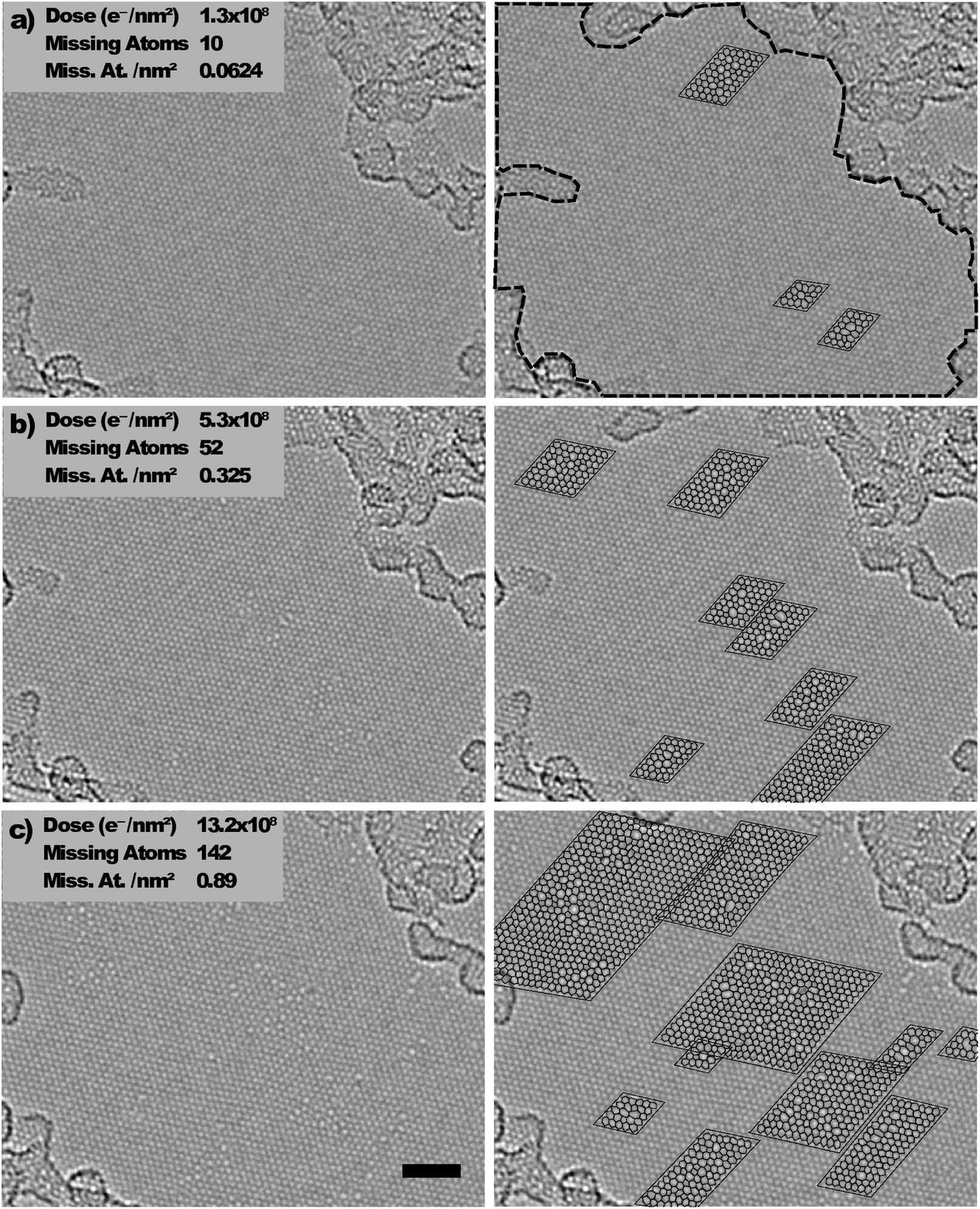}

\caption{Multi-vacancy defects with increasing dose under 100~keV observation
of $^{12}\textrm{C}$ graphene (a-c). Images are shown without (left)
and with overlay of the atomic configuration (right). The dashed line
in (a) indicates the area that is used to calculate the missing atoms
per area. The scale bar is 2~nm.\label{fig:examplesequence}}

\end{figure}

Here, we present an extensive measurement and analysis of electron-beam
induced displacements. We directly count the number of ejected atoms
under irradiation as functions of dose, dose rate, and electron energy.
Suspended single layer graphene sheets provide the perfect test sample
for this analysis: They can be prepared in a precisely defined geometry
(1 atomic layer thick, hexagonal lattice, with practically no defects
initially), are relatively easy to model, and the number of ejected
atoms in multi-vacancy configurations can be directly obtained from
HRTEM images \cite{Meyer2008a,Kotakoski2011}. Under 80~keV electron
irradiation, the defect free graphene lattice remains undisturbed
up to very high doses \cite{Meyer2008a,Meyer2011} but knock-on damage
begins already a few keV above this energy \cite{LUCAS1964,Crespi1996a,Banhart1999,Smith2001,Krasheninnikov2005,Zobelli2007a,Molhave2007,Warner2009a}.
Importantly, for energies near the knock-on threshold, the changes
in the lattice occur slowly, so that the appearance and growth of
multi-vacancies can be directly observed in real time. In this way,
we can count the number of lost atoms as a function of applied dose
and for different acceleration voltages, hence providing a direct
measurement of the knock-on cross section.

We present insights from a tremendous data set that was obtained for
the purpose of quantitating the radiation damage in graphene: We have
obtained and analyzed image sequences as shown in Fig. 1 for many
acceleration voltages (80, 90, 95, 100 kV), and for both, the $^{12}\textrm{C}$
{}``normal'' graphene sample and isotope-enriched $^{13}\textrm{C}$
{}``heavy graphene'' samples. For all of this data, the defect configurations
were analyzed at different doses of exposure, and the number of missing
atoms was counted (see supplementary information for further examples
from the data set). We also studied $^{12}\textrm{C}$ graphene under
20 keV electron irradiation \cite{Kaiser2011a}, in order to obtain
a further distinction between knock-on damage and other effects such
as chemical etching or radiolysis. 

Experimentally, we prepared graphene membranes by mechanical exfoliation
and transfer to TEM grids as described previously \cite{Meyer2008b},
and by chemical vapor deposition (CVD) followed by transfer to TEM
grids, as described in Ref. \cite{Park2010}. We assume that these
samples contain the natural isotope composition in carbon, which is
98.9\% $^{12}\textrm{C}$ and 1.1\% $^{13}\textrm{C}$. In addition,
we synthesized {}``heavy graphene'' samples made from $^{13}\textrm{C}$,
by CVD. The synthesis recipe for the $^{13}\textrm{C}$ graphene followed
the same procedure as we described in Ref. \cite{Park2010}, except
that the standard methane precursor was replaced by 99\% $^{13}\textrm{C}$
enriched methane (purchased from Sigma-Aldrich). We aligned an image
side aberration-corrected FEI Titan 80-300 for HRTEM imaging at 80,
90, 95 and 100 kV, hence providing a closely spaced series around
the threshold voltage \cite{Smith2001}. The spherical aberration
was set to ca. $20\,\mu\textrm{m}$ and images were recorded at Scherzer
defocus. Under these conditions, dark contrast can be directly interpreted
in terms of the atomic structure. For 20~kV imaging, we used an image-side
aberration corrected Zeiss Libra as described in Ref. \cite{Kaiser2011a}.
In all experiments, long image sequences of the graphene samples were
recorded (see supplementary videos), typically consisting of $\sim100$
images with 1~s exposures recorded at 2-4~s intervals and typical
dose rates of $10^{6}\,\frac{e^{-}}{\textrm{nm}^{2}\cdot s}$. The
sample is under continuous irradiation, only the beam shutter behind
the sample is used. We analyze the creation and the increase in the
density of vacancies and multi-vacancies in the image sequences. This
approach was feasible up to ca. 100~kV, while at 120~kV the damage
occured too quickly compared to the time or dose needed to acquire
an HRTEM image with a sufficient signal to noise ratio. 

\begin{figure}
\includegraphics[width=0.78\columnwidth]{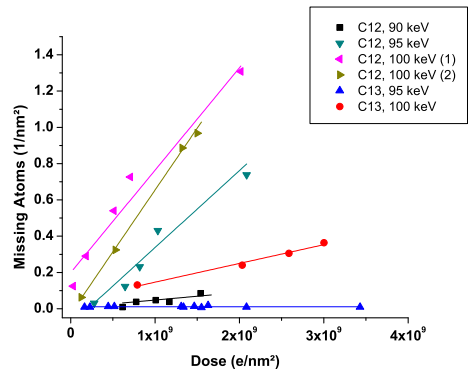}

\caption{Number of displaced atoms vs. dose and electron energy. For the 100~keV
case, two different dose rates were compared, with (2) having a ca.
$3\times$ higher dose rate than (1).\label{fig:displacements}}

\end{figure}

Example images (for 100~kV, $^{12}\textrm{C}$) are shown in Fig.
1.: The left hand side shows images from an image sequence recorded
at 100~kV and the right hand side shows the same images with a structure
overlay of the atomic configuration. The analysis of such atomic configurations
has been described in more detail previously \cite{Kotakoski2011,Meyer2008a}:
the multi-vacancies reconstruct into configurations that involve primarily
carbon pentagons, heptagons and octagons as well as other non-hexagonal
rings, and can be well assigned from HRTEM data. For counting the
atoms, we draw a supercell around the defect clusters, such that the
boundary of this cell does not intersect any defect (see Fig. 1).
Moreover, the supercell must not contain unpaired dislocation cores,
which can be easily verified by counting the number of unit cells
on opposing sides of the parallelogram. We then calculate the number
of atoms that should be within this cell for the defect free case,
and compare it to the number of atoms actually present. 

The results of this assessment are shown in Fig. 2, where the number
of lost atoms vs. total dose per area is shown for all experiments.
A linear fit is made for each case, and the slope directly provides
the experimental knock-on cross section (since a small initial damage
may be created before the first image is recorded some of the lines
do not go through the origin). Two independent measurements were made
for the 100kV case, but with a $3\times$ different \emph{dose rate
}($3.5\cdot10^{5}\frac{e^{-}}{\textrm{nm}^{2}\textrm{s}}$ and $1\cdot10^{6}\,\frac{e^{-}}{\textrm{nm}^{2}\textrm{s}}$).
From the nearly identical result, we can exclude a dose rate effect
within our experimental precision. Under 80~keV irradiation (not
shown in Fig. 2), no vacancies were formed in pristine areas up to
very high doses (beyond $10^{10}\,\frac{e^{-}}{\textrm{nm}^{2}}$). 

\begin{figure}
\includegraphics[width=1\columnwidth]{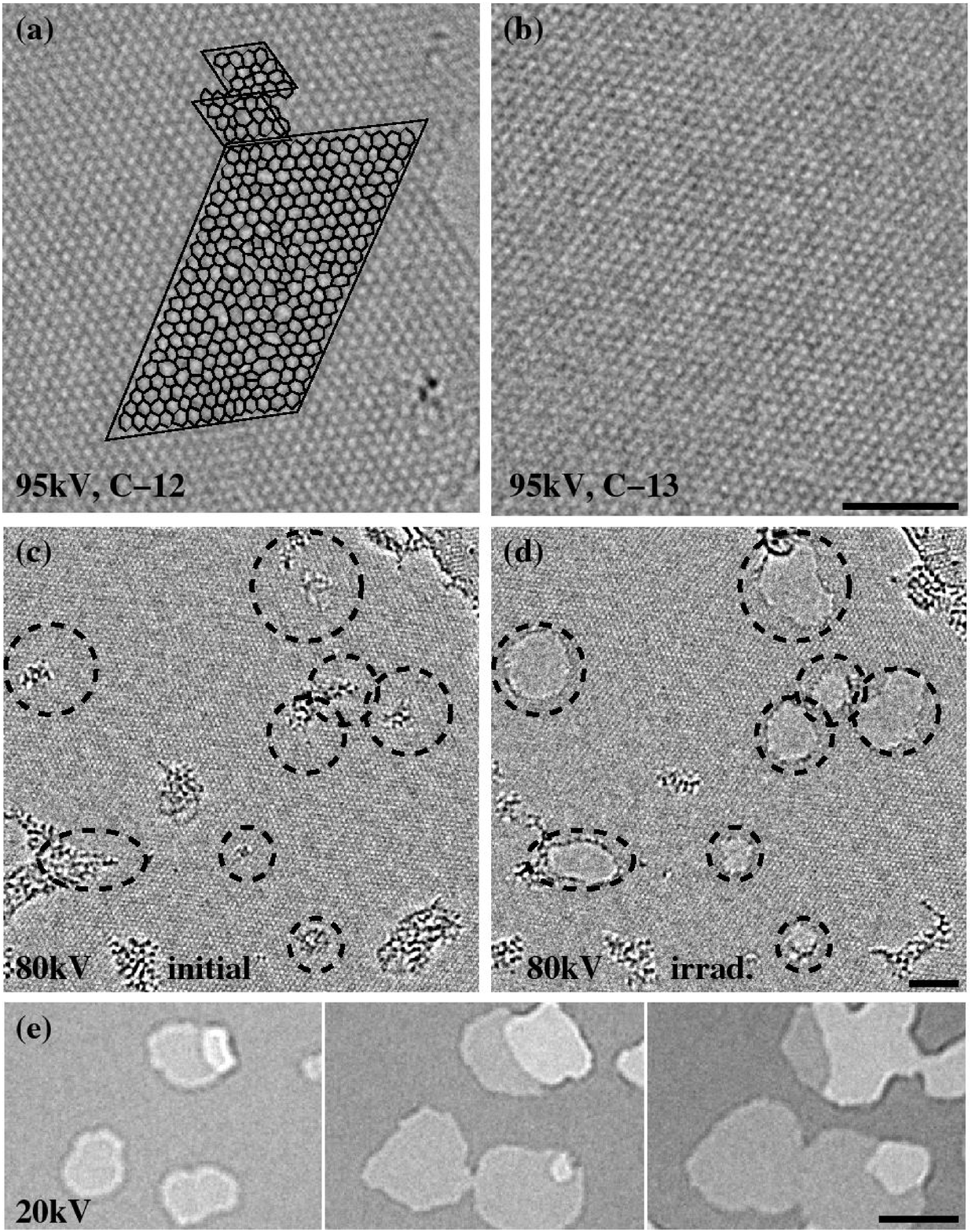}

\caption{Atomic displacements (knock-on damage) vs. chemical etching. The comparison
between $^{12}\textrm{C}$ and $^{13}\textrm{C}$ graphene shows that
the formation of vacancies within the pristine lattice is a direct
knock-on damage effect (a+b, $^{12}\textrm{C}$ and $^{13}\textrm{C}$
sample after a dose of $1.4\cdot10^{9}\,\frac{e^{-}}{\textrm{nm}^{2}}$
at 95~keV). In contrast, the formation of extended holes in graphene
is always induced by contamination on the sheet (c, initial image,
and d after exposure to ca. $10^{9}\,\frac{e^{-}}{\textrm{nm}^{2}}$
), and the damage rate depends on the vacuum levels. Dashed circles
in c and d denote same areas. (e) Image sequence showing the growth
of holes in graphene at 20~keV (example shown for a bi-layer area).
Scale bars are 2~nm (a-d) and 5nm (e).\label{fig:c12c13andetching}}

\end{figure}

We begin our discussion by pointing out the clear difference between
the $^{12}\textrm{C}$ and $^{13}\textrm{C}$ graphene membranes,
and the differences between knock-on damage and a chemical etching
effect. We find that the generation of vacancies within initially
pristine, clean and defect free graphene membranes depends on the
acceleration voltage, and also on the isotope composition ($^{12}\textrm{C}$
vs. $^{13}\textrm{C}$). Fig. 3a,b shows graphene membranes of the
two isotopes after exposure to 95~keV electrons, where the difference
is most clearly visible. Hence, this must be a result of a direct
collision between a beam electron and the carbon nucleus: Any chemical
effect, or ionization damage, would not distinguish between $^{12}\textrm{C}$
and $^{13}\textrm{C}$. As a side remark, we note that graphene membranes
made from $^{13}\textrm{C}$ might provide an even better TEM sample
support than ordinary single-layer graphene: The contrast background
in HRTEM is identical, but the radiation damage rate is lower.

However, in contrast to the vacancy formation, the growth of extended
holes in graphene \cite{Girit2009} is not predominantly a knock-on
damage effect: We found that the growth rate of holes in graphene
only weakly depends on the electron energy on a wide range of 20~keV
to 100~keV: Holes still form and grow in graphene under 20 keV irradiation
(Fig. 1c-e), and may even grow faster at low voltages (see supplementary
information). This is in stark contrast to expectations from knock-on
damage, where the threshold for displacing edge atoms is expected
to be near 50~keV \cite{Kotakoski2011c}. As shown in Fig. \ref{fig:c12c13andetching}c-d,
the extended holes always nucleate at contamination sites. We noticed
that their growth rate is related to the vacuum levels, which varied
in the range of $10^{-6}$ to $10^{-7}$ millibar, e.g. with use of
the cold trap in the column, the time after insertion of the sample,
or different outgassing rates of different sample holders. We conclude
that this is beam-induced etching with residual water or oxygen in
the system as described in Refs. \cite{Molhave2007,Barreiro2012}.
Therefore, we count the formation of vacancies in initially clean
and defect free areas as knock-on effect, but do not take into account
the extended holes that nucleate at contamination sites.

\begin{figure}
\includegraphics[width=0.88\columnwidth]{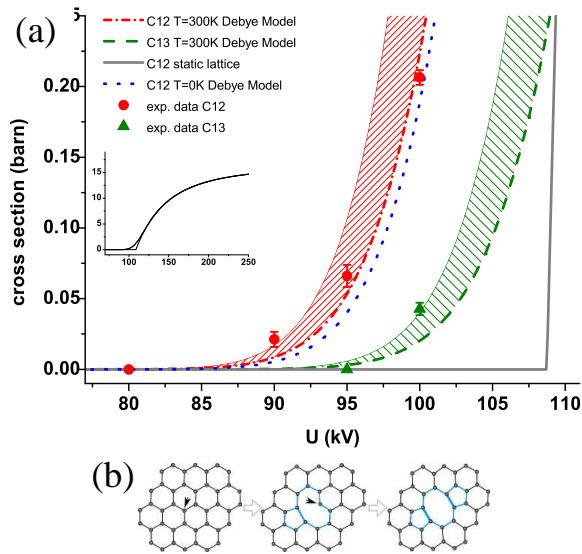}

\caption{(a) Measured and calculated knock-on displacement cross sections.
The lower boundary of the shaded areas correspond to the calculated
cross section, while the upper boundary is twice the calculated value
(as would be expected for correlated sputtering). Inset shows the
calculations for $^{12}\textrm{C}$, 300K and static lattice on a
larger energy range. (b) Correlated displacement of carbon atoms.
After creation of a mono-vacancy, one carbon atom remains with a dangling
bond and a much lower emission threshold. Subsequent sputtering of
this atom may effectively double the cross section.}

\end{figure}

The analysis of our results culminates in the plot shown in Fig. 4.
Here, each of the slopes from Fig. 2 provides one data point for a
measured displacement cross section. The error bars indicate the statistical
variation (standard deviation) in the data. Also shown in Fig. 4 are
calculated curves from existing and new calculations that will be
discussed below. For two curves, we show a shaded area between $1\times$
and $2\times$ the calculated cross section, since correlated sputtering
of carbon atoms may increase the observed atom loss by up to a factor
of two: After creation of a mono-vacancy by electron impact, one carbon
atom is left with a dangling bond and a much lower emission threshold
\cite{Krasheninnikov2005,Kotakoski2011c}. Subsequent (and much more
rapid) sputtering of this atom may effectively double the rate of
atom loss (Fig. 4b). As a competing mechanism, two mono-vacancies
that are created close to each other may combine and form a stable
di-vacancy (since impacts of energetic electrons can rotate bonds
in graphene \cite{Kotakoski2011,Kotakoski2011a}, it is likely that
exposure to the beam increases also the diffusivity of vacancies).
In this case, the sputtering rate would not have to be doubled. Qualitatively,
one would expect that correlated sputtering is dominant close to the
threshold, while nearby mono-vacancies are more likely generated at
higher electron energies (this is further discussed in the supplement).
In any case, the resulting multi-vacancy configurations contain only
very few undercoordinated carbon atoms, while the 3-coordinated atoms
in the reconstructed configurations are expected to have an emission
threshold similar to that of an atom in the pristine graphene sheet
\cite{Kotakoski2011a}. Another conceivable mechanism that might have
an influence on the experimental results, namely the annealing of
vacancies with mobile carbon adatoms, can not be dominant in our experiment
as evidenced by the absence of a dose rate effect. Hence, we expect
a rate of atom loss in-between $1\times$ and $2\times$ of the value
calculated for pristine graphene. 

The cross section for Coloumb scattering between an electron and a
corresponding target nucleus was derived by Mott \cite{Mott1929}.
McKinley and Feshbach have found an analytic expression for the Mott
scattering cross section as a function of the maximum transferred
energy \cite{W.McKinley},

\begin{equation}
\begin{array}{ccc}
\sigma_{D} & = & \frac{4Z^{2}E_{R}^{2}}{m_{e}^{2}c^{4}}(\frac{T_{max}}{T_{thr}})\pi a_{0}^{2}(\frac{1-\beta^{2}}{\beta^{4}})\{1+2\pi\alpha\text{\ensuremath{\beta\sqrt{\frac{T_{thr}}{T_{max}}}}}\\
 &  & -\frac{T_{thr}}{T_{max}}[1+2\pi\alpha\beta+(\beta^{2}+\pi\alpha\beta)ln(\frac{T_{max}}{T_{thr}})]\}\end{array}\label{eq:1-1}\end{equation}

where $Z$ is the atomic number of the target atoms, $E_{R}=13.6eV$
the Rydberg energy, $a_{0}=5.3\cdot10^{-11}\,\textrm{m}$ the Bohr
radius of the hydrogen atom, $\beta=\frac{v_{e}}{c}$ (electron velocity
$v_{e}$ divided by the speed of light $c$), $m_{e}$ the mass of
th electron and $\alpha\approx\frac{Z}{137}$. $T_{max}$ represents
the maximum transferred energy in the collision event and $T_{thr}$
a threshold energy for atomic displacement. Without modifications,
equation (\ref{eq:1-1}) is suitable to evaluate the total {}``knock-on''
cross section for an atom at rest with a given ejection threshold
energy. The curve from equation (\ref{eq:1-1}) ({}``static lattice''
in Fig. 4), features a rather sharp onset of radiation damage with
increasing acceleration voltage: The cross section is zero up to a
well defined threshold (here 108 kV), and then rises to several barn
(beyond all our measured values) only a few kV above this threshold.
Changing the displacement threshold in the Mc-Kinley Feshbach formula
predominantly shifts this curve sideways, but does not affect the
sharp onset. Hence, independent of the free parameter $T_{thr}$ this
approximation is in clear contrast to our experiment, which shows
a smooth onset of the damage cross section between 80 and 100~keV.

Remarkably, our data can be explained by considering the effect of
the struck atom's vibrations on its own displacement. While the effect
has been discussed earlier \cite{Brown1959,Iwata,Zag1983}, our measurement
provides precise experimental evidence of this intriguing effect.
In essence, it means that an atom that is struck by an electron while
it happens to move parallel to the electron beam can obtain a higher
maximum transferred energy $T_{max}$ than if it were static. For
our calculation, we approximate the phonon distribution of the material
in the framework of the Debye model. We use the Debye temperature
calculated for out of plane vibrations in graphene of $\theta_{D}=1287\textrm{K}$
from Ref. \cite{Tewary2009}. Since $\theta_{D}$ depends on the speed
of sound, it follows for the Debye temperature of $^{13}\textrm{C}$
that $\theta_{D}^{13}=\sqrt{\frac{12}{13}}\theta_{D}^{12}$. We extract
the distribution of atom velocities in the beam direction from the
model, calculate the maximum transferred energy $T_{max}(v,E)$ as
a function of the atom velocity $v$ and electron energy $E$, and
obtain the weighted sum of the sputtering cross section numerically
(see supplementary information). In other words, we still use the
Mott scattering cross section, but we consider that the atom is not
at rest initially. The threshold energy of $T_{thr}=22\,\textrm{eV}$
was taken from first principles calculations \cite{Kotakoski2010}
without any adjustments (Refs. \cite{Zobelli2007a,Krasheninnikov2005}
give similar values). With this value, the smooth onset of knock on
damage between 80-100~keV is very well reproduced. For the first
time, no adjustment to the calculated threshold energy $T_{thr}$
is needed to explain the data, as was the case in previous studies
\cite{Zobelli2007a,Kotakoski2010}: Remarkably, the previous mismatch
between theory and experiments was not due to inadequate calculations
of $T_{thr}$, but because the effects of lattice vibrations on the
elastic collision were not considered. Interestingly, the calculated
curves are almost identical for the zero-kelvin and room-temperature
case (Fig. 4, $^{12}\textrm{C}$ T=0~K and T=300~K curves). This
implies that already the zero-point energy of the phonon modes is
sufficient to explain the increased sputtering cross section as compared
to the static lattice.

In summary, we have made an accurate measurement of atomic displacement
cross sections for carbon atoms in single-layer graphene. The cross
section smoothly rises from practically zero ($10^{-4}$ barn) at
80~keV to $\sim0.2$~barn at 100~keV. In practice this means that
80~keV imaging of defect free graphene is easily possible, while
already 100~keV TEM images might not represent the original configuration
of a sample. A static lattice model is not sufficient to model the
process, and the contribution of atomic motion adds significantly
to knock-on damage cross sections near the threshold. The difference
between $^{12}\textrm{C}$ and $^{13}\textrm{C}$ isotopes is detectable
and further confirms the model. While the results on graphene will
be important for HRTEM studies of this material and related ones (especially
carbon nanotubes), the generalized insights to radiation damage mechanisms
should be more generally applicable to any material where knock-on
damage is important. Our results show that knock-on displacement cross
sections can be modeled with high accuracy, if lattice vibrations
are taken into account. 

\textbf{Acknowledgments:} We acknowledge financial support by the
German Research Foundation (DFG) and the Ministry of Science, Research
and the Arts (MWK) of the state Baden-Württemberg within the Sub-Angstrom
Low-Voltage Electron Microscopy project (SALVE). A.V.K. and J.K. acknowledge
the support from the Academy of Finland through several projects.
V.S. acknowledges EC grant No 266391 related to the project ELECTROGRAPH
(FP7/2007-2013). S.R. acknowledges support by Korea WCU R32-2008-000-10082-0.

\part*{Supplementary information}

\section{Supplementary experimental data}

Figures 5-7 show additional image sequences and the analysis of defect
geometries with the number of missing atoms. Figs. 5 and 6 show the
radiation damage in $^{12}\textrm{C}$ graphene under 90~keV and
95~keV electron irradiation. For Fig. 6, it should be noted that
the extended hole that grows on the right-hand side of the image was
not taken into account (as discussed above, these holes nucleate at
contamination sites and grow via a beam-induced etching effect). The
sequence in Fig. 6 also shows two of the rare cases where a heavier
contamination atom is trapped in a vacancy (these are also counted
as a lost carbon atom). Fig. 7 shows the $^{13}\textrm{C}$ graphene
sample under 100~keV electron irradiation. The rate of atom loss
in this sample is significantly reduced compared to the $^{12}\textrm{C}$
sample (compare Fig. 1 in the main article). \\
\\
\\

\begin{figure}
\includegraphics[width=1\columnwidth]{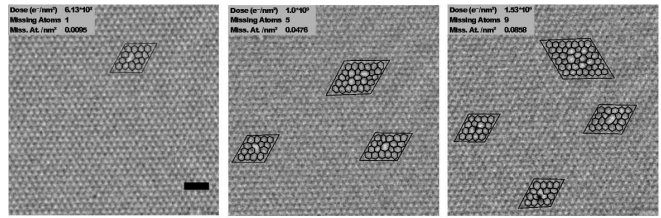}

\caption{Radiation damage in $^{12}\textrm{C}$ graphene under 90~keV electron
irradiation. Scale bar 1nm.}

\end{figure}

\begin{figure}
\includegraphics[width=1\columnwidth]{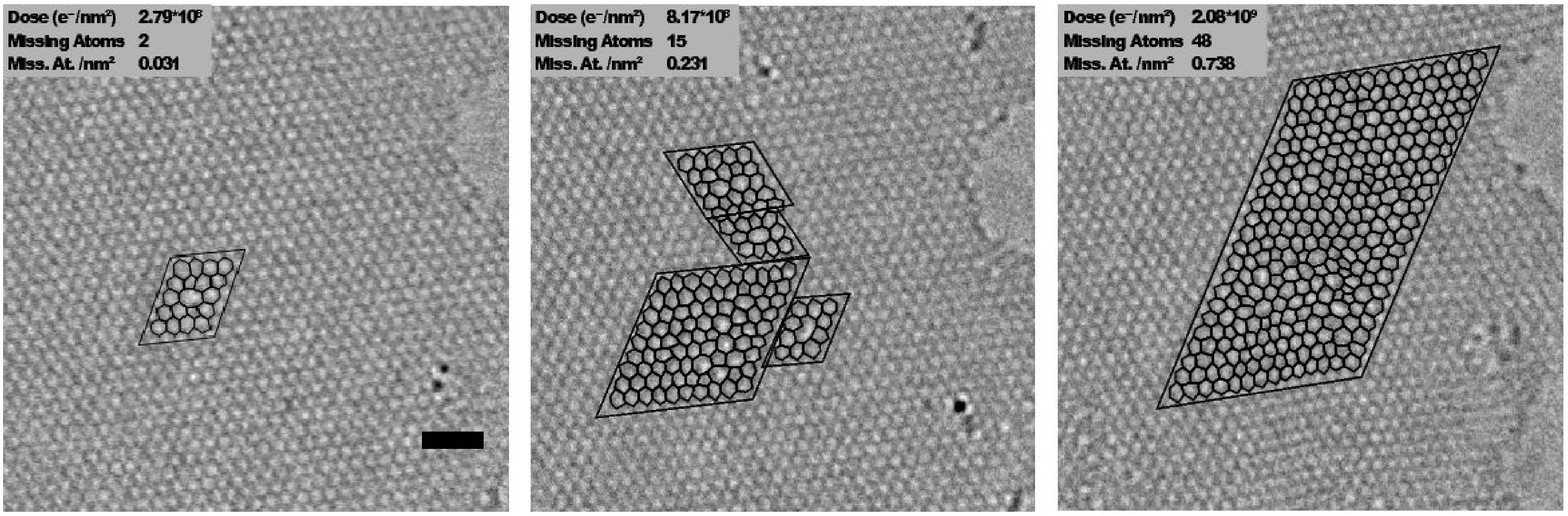}

\caption{Radiation damage in $^{12}\textrm{C}$ graphene under 95~keV electron
irradiation. Scale bar 1nm}

\end{figure}

\begin{figure}
\includegraphics[width=1\columnwidth]{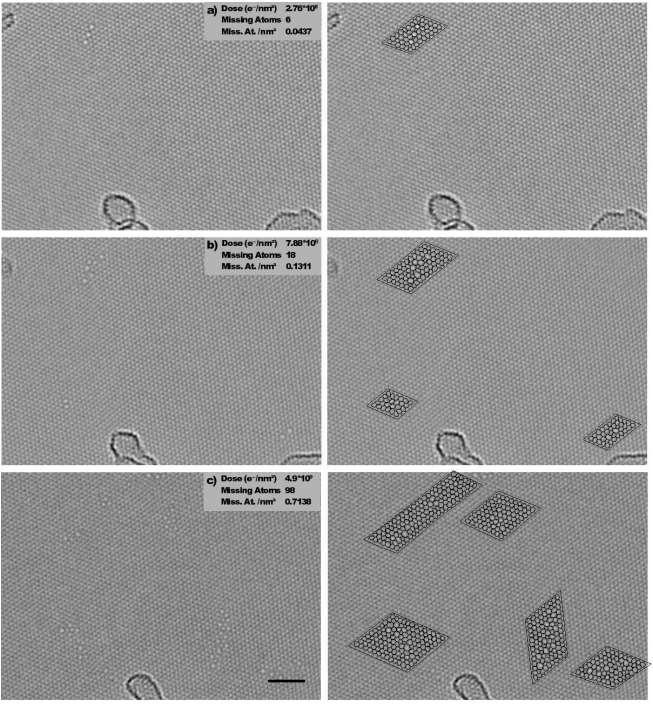}

\caption{Radiation damage in $^{13}\textrm{C}$ graphene under 100~keV electron
irradiation. Scale bar 2~nm}

\end{figure}

\section{Details of the displacement cross section calculation}

The cross section for Coloumb scattering between an electron and a
corresponding target nucleus has been derived by Mott in \cite{Mott1929}.
McKinley and Feshbach have found an analytic expression for the Mott
scattering cross section as a function of the maximum transferred
energy \cite{W.McKinley}.\begin{equation}
\begin{array}{ccc}
\sigma_{D} & = & \frac{4Z^{2}E_{R}^{2}}{m_{e}^{2}c^{4}}(\frac{T_{max}}{T_{thr}})\pi a_{0}^{2}(\frac{1-\beta^{2}}{\beta^{4}})\{1+2\pi\alpha\text{\ensuremath{\beta}}\sqrt{\frac{T_{thr}}{T_{max}}}\\
 &  & -\frac{T_{thr}}{T_{max}}[1+2\pi\alpha\beta+(\beta^{2}+\pi\alpha\beta)\ln(\frac{T_{max}}{T_{thr}})]\}\end{array}\label{eq:1-1}\end{equation}
where $Z$ is the atomic number of the target atoms, $E_{R}=13.6eV$
the Rydberg energy, $a_{0}=5.3*10^{-11}m$ the Bohr radius of the
hydrogen atom, $c$ the speed of light, $\beta=\frac{v_{e}}{c}$ ($v_{e}$
velocity of the electron), $c$ the speed of light, $m_{e}$ the mass
of the electron, and $\alpha=\frac{Z}{137}$. $T_{max}$ represents
the maximum transferred energy in the collision event and $T_{thr}$
a threshold energy for atomic displacement. For an atom at rest, the
maximum transferred energy in an elastic impact of the beam electron
is given by expression

\begin{equation}
T_{max}(E)=\frac{2E(E+2m_{e}c^{2})}{c^{2}m_{n}}\label{eq:tmaxstd}\end{equation}
where $E=eU$ is the kinetic energy of the electrons and $m_{n}$
the mass of the atomic nucleus. This expression is based on the (well
justified) approximations that the atomic nucleus is much heavier
than the electron ($m_{n}\gg m_{e}$), and that its rest energy is
much larger than the beam energy ($c^{2}m_{n}\gg E$).

The above expression is commonly used to calculate sputtering cross
sections and thresholds. Our modified calculation now explicitly takes
into account the fact that the target atom is not at rest due to the
vibrations of the lattice. The importance of lattice vibrations as
well as thermal effects for the displacement of atoms, irradiated
with energetic particles was first suggested in 1959 by Brown and
Augustyniak and subsequently mentioned in several publications \cite{Brown1959,Iwata,Zag1983}.
The effect is not obvious, because compared to the energy that is
needed to displace an atom, its vibrational energy is very small.
However, the fact that the atom is moving, significantly changes the
transferred momentum and energy upon impact of the fast electron.
By considering again a (relativistic) elastic collision between the
beam electron and a moving target atom, the maximum transferred energy
can be written as

\begin{equation}
\tilde{T}_{max}(v,E)=\frac{r*(r+\frac{2t}{c})+\frac{t\text{\texttwosuperior}}{c\text{\texttwosuperior}}}{2m_{n}}\label{eq:tmaxmod}\end{equation}
with: $r=\frac{1}{c}\sqrt{E(E+2m_{e}c\text{\texttwosuperior})}+m_{n}v$ 

and $t=\sqrt{(E+E_{n})(E+2m_{e}c\text{\texttwosuperior}+E_{n})}$
\\
where $v$ is the initial velocity and $E_{n}=\frac{m_{n}v\text{\texttwosuperior}}{2}$
is the initial kinetic energy of the target atom. Here, $v$ refers
only to the component of the velocity that is normal to the graphene
plane, i.e. parallel to the electron beam.

We now consider how the cross section (eq. \ref{eq:1-1}) is changed
for the case of a moving target. The analytical expression for the
original Mott Series \cite{Mott1929,Mott1932} (which is accurate
up to the medium-Z elements \cite{McKinley1948}) can be written as
\cite{McKinley1948}

\begin{equation}
\sigma(\theta)=\sigma_{R}(\theta)[1-\beta^{2}\sin^{2}(\frac{\theta}{2})+\pi*\frac{Ze^{2}}{\hbar c}\beta\sin(\frac{\theta}{2})(1-\sin(\frac{\theta}{2}))]\label{eq:mott cross section}\end{equation}
with $\beta=\frac{v}{c}$, $\theta$ being the scattering angle and
$\sigma_{R}$ the classical Rutherford scattering cross section:\[
\sigma_{R}(\theta)=(\frac{Ze^{2}}{4\pi\epsilon_{0}2m_{0}c^{2}})^{2}\frac{1-\beta^{2}}{\beta^{4}}\csc^{4}(\frac{\theta}{2}).\]

If the target atom is at rest, the angular dependence of the transferred
energy can be described as

\begin{equation}
T(\theta)=T_{max}(E)\sin^{2}(\frac{\theta}{2})\label{eq:winkel abhängigkeit von tmax}\end{equation}
(with $T_{max}$ as given in eq. \ref{eq:tmaxstd}) and then integration
over the scattering angles yields equation \ref{eq:1-1}.

For a moving target, the angular dependence of the transferred energy
(with the atom initially moving) can be written as:

\begin{equation}
\tilde{T}(\theta)=\frac{r*(r-\frac{2t}{c}*\cos(\theta))+\frac{t^{2}}{c^{2}}}{2m_{n}}\label{eq:tmax_new}\end{equation}
with $r$ and $t$ as given above. This can be rewritten and approximated
as \begin{equation}
\tilde{T}(\theta)=\tilde{T}_{max}(v,E)*(\sin^{2}(\frac{\theta}{2})(1-x)+x)\thickapprox\tilde{T}_{max}*\sin^{2}\text{(\ensuremath{\frac{\theta}{2}}) }\label{eq:Tmax approx with sinus}\end{equation}
with $x=1-\frac{4*r*\frac{t}{c}}{(r+\frac{t}{c})\text{\texttwosuperior}}$,
where the approximation is valid for $x\ll1$. 

Hence, we obtain the same expression for $\sigma_{D}$ as in equation
(\ref{eq:1-1}), except that $T_{max}(E)$ is replaced by the new
$\tilde{T}_{max}(v,E)$: For the case of a moving target, we obtain

\begin{equation}
\begin{array}{c}
\tilde{\sigma}_{D}(v,E)=\sigma_{D}(\tilde{T}_{max}(v,E))=\,\,\,\,\,\,\,\,\,\,\,\,\,\,\,\,\,\,\,\,\,\,\,\,\,\,\,\,\,\,\,\,\,\,\,\,\,\,\,\,\,\,\,\,\,\\
\frac{4Z^{2}E_{R}^{2}}{m_{e}^{2}c^{4}}(\frac{\tilde{T}_{max}}{T_{thr}})\pi a_{0}^{2}(\frac{1-\beta^{2}}{\beta^{4}})\{1+2\pi\alpha\text{\ensuremath{\beta}}\sqrt{\frac{T_{thr}}{\tilde{T}_{max}}}\\
-\frac{T_{thr}}{\tilde{T}_{max}}[1+2\pi\alpha\beta+(\beta^{2}+\pi\alpha\beta)\ln(\frac{\tilde{T}_{max}}{T_{thr}})]\}.\end{array}\label{eq:newformula1}\end{equation}

For our parameters (i.e. acceleration voltages, Debye temperature,
atomic mass) we obtain $x\approx0.003$, meaning that the approximation
is well justified. The approximation can also be understood qualitatively,
as that if the electron velocity is much higher than the atom velocity,
the scattering angle $\theta$ is not significantly changed by the
motion of the atom. Only for very low electron energies (on the order
of 10~keV or below), one would need to consider a modification to
equation \ref{eq:1-1} or \ref{eq:newformula1}.

The mean square velocity of an atom can be calculated within the framework
of the Debye model to be \cite{Sinnemann1992}\begin{equation}
\overline{v\text{\texttwosuperior}}=\frac{9k_{b}}{8m_{n}}\theta_{D}+\frac{9k_{b}T}{m_{n}}(\frac{T}{\theta_{D}})\text{\textthreesuperior}\intop_{0}^{\frac{\theta_{D}}{T}}\frac{x\text{\textthreesuperior}}{\exp(x)-1}dx\end{equation}
and it is Gaussian distributed (central limit theorem). Here, $k_{b}$
is the Boltzmann constant, $T$ the temperature, and $\theta_{D}$
the Debye temperature. The Debye model naturally contains a Bose-distribution
for the phonon modes that also includes the quantum mechanical zero
point motion for all vibration modes; meaning that the atoms are not
at rest even at zero temperature. The normalized Gaussian distribution
yields a probability for finding an atom with a velocity between $v$
and $v+dv$: \begin{equation}
P(v,T)dv=\frac{1}{\sqrt{\pi\overline{v^{2}}}}\exp(\frac{-v^{2}}{\overline{v^{2}}})dv.\label{eq:probabiooty}\end{equation}

Now, that we have the probability distribution for the velocities
(\ref{eq:probabiooty}) and using the scattering cross section (\ref{eq:newformula1})
with the velocity-dependent expression for $T_{max}(v)$ (\ref{eq:tmaxmod}),
we can integrate over all velocities weighted by their probability
in order to obtain the total cross section:

\begin{equation}
\sigma(T,E)=\intop_{-\infty}^{\infty}P(v,T)\sigma_{D}(\tilde{T}_{max}(v,E))\Theta(\tilde{T}_{max}(v,E)-T_{thr})dv.\label{eq: sputtering cross section}\end{equation}

This last integration step was done numerically. The heaviside step
function $\Theta(\tilde{T}_{max}(v,E)-T_{thr})$ ensures that only
positive values for $\sigma_{D}$ are summed up (equations \ref{eq:1-1}
or \ref{eq:newformula1} give non-physical, negative values if $T_{max}<T_{thr}$,
which must be replaced by a zero scattering probability). The numerical
integration was done over a range $-8\sqrt{\overline{v^{2}}}\leq v\leq8\sqrt{\overline{v^{2}}}$,
providing the curves as shown in Fig. 4 of the main article.

\section{Correlated sputtering}

Here, we briefly discuss the two competing mechanisms of correlated
sputtering and the merging of nearby vacancies. After removing one
carbon atom from the graphene lattice, in principle there are three
atoms in an undercoordinated configuration. However, the configuration
reconstructs to incorporate a carbon pentagon, which closes two of
the open bonds and leaves only one atom undercoordinated \cite{Banhart2011}.
The displacement threshold for this atom is only 14.7~eV \cite{Kotakoski2011c},
compared to 22~eV for an atom in the pristine lattice. Our HRTEM
observations show predominantly di-vacancies and (clustered) multi-divacancies,
and exceptionally few single vacancies or other undercoordinated atoms.
\begin{figure}
\includegraphics[width=0.95\columnwidth]{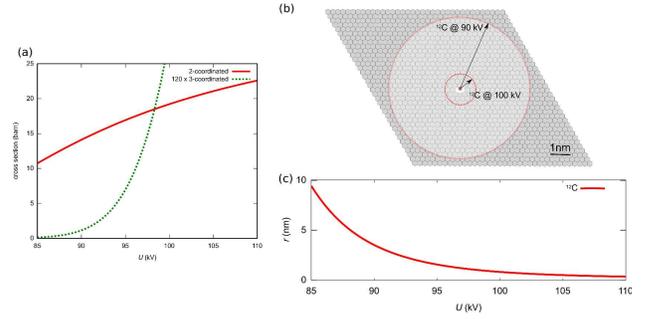}

\caption{(a) Cross section for sputtering a 2-coordinated carbon atom compared
to 120x the cross section of sputtering a 3-coordinated carbon (which
describes the probability of sputtering a 3-coordinated carbon within
a 1nm radius around the vacancy). (b+c) Radius over which sputtering
the undercoordinated atom would be equally likely as sputtering an
additional carbon from the pristine lattice. (b) Graphic illustration
for 90 and 100kV; (c) plot of the radius vs. electron energy.}

\end{figure}

Two scenarios appear reasonable in the light of these observations.
First, one might expect that the undercoordinated carbon atom is sputtered
shortly after the initial vacancy is generated. In this case, the
observed rate of atom loss would be twice the rate that is calculated
for the pristine graphene lattice. Second, mono-vacancies that form
in close proximity to each other may cluster and form di-vacancies
without loosing additional atoms. As a starting point for quantifying
these effects, we now compare (a) the probability (cross section)
for sputtering the undercoordinated carbon atom, and (b) the probability
for forming two mono-vacancies in close proximity. Fig. 8a shows the
probabilites (as cross sections) for sputtering one undercoordinated
atom compared to sputtering one carbon atom within a 1nm radius (120
atoms) of the original defect. From the graph it is clear that correlated
sputtering will be dominant at lower electron energies and less significant
at high energies, but quantification is difficult: Although clustering
of defects is clearly observed, it is not clear at which rate or over
which distances it occurs (the choice of 1nm as the example given
above and in Fig. 8a is arbitrary). Fig 8b+c shows the analysis represented
in a different way: It describes a radius over which sputtering an
additional carbon atom from the pristine lattice would be equally
likely as sputtering the undercoordinated atom at the vacancy. In
our data (Fig. 4 of the main article), it indeed appears that the
lowest energy data points ($^{12}\textrm{C}$ at 90kV, $^{13}\textrm{C}$
at 100kV) correspond to correlated sputtering, while $^{12}\textrm{C}$
at 100kV does not. This is in agreement with the above discussion
if we assume that mono-vacancies can cluster into divacancies over
a distance of 1nm or less. In any case, we consider the range from
1x to 2x of the calculated cross section in the pristine lattice as
a (material-specific) uncertainty in the model.

\section{Chemical etching}

In this section we expand our discussion of chemical etching in graphene
under TEM imaging conditions. Although the focus of our analysis is
on knock-on damage effects, it is also important to identify and separate
other effects that are not knock-on damage. From a large number of
experiments on graphene in the TEM, we conclude that the growth of
extended holes is not predominantly an effect of knock on damage.
This is particularly surprising since, under 80 keV irradiation, direct
sputtering of edge atoms must also occur. However, the experiments
show that it can not be the dominant mechanism.

Fig. 9a-c shows the typical formation and evolution of a hole in a
graphene membrane under 80~keV electron irradiation. Initially, a
contamination site is present (which may be attached to a defect site).
Electron-energy loss spectroscopy and energy-dispersive x-ray spectroscopy
indicated that silicon is a predominant contamination in our samples,
and no other elements were detected. However, it can not be excluded
that other reactive species are present in smaller quantities. In
any case, we almost exclusively observe the formation of holes at
sites with heavier (than carbon) contamination (Fig. 9a). The hole
rapidly grows while the contamination is present and attached to the
open edges (Fig. 9b), but eventually a hole with pristine (carbon-only)
edges is formed (Fig. 9c). Such a hole can now be observed over extended
times and up to high doses \cite{Girit2009}. It slowly grows under
the beam. We have quantified the hole growth by measuring the hole
perimeter as a function of dose, shown for part of our data in Fig.
9d. Hole formation and growth was observed in the same way also under
20kV HRTEM imaging \cite{Kaiser2011a}.

\begin{figure}
\includegraphics[width=1\linewidth]{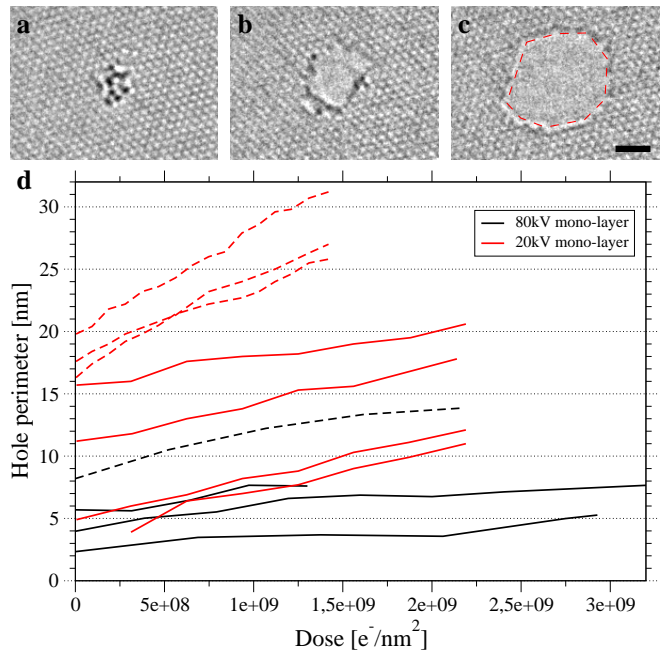}

\caption{Analysis of chemical etching. (a-c) Typical sequence of hole formation
in graphene (scale bar 1nm). (a) The holes nucleate at a contamination
site. (b) The hole rapidly grows while heavier contamination atoms
are still bound to the open edges. (c) Eventually a {}``clean''
hole in graphene is formed (only carbon atoms at the edges), and hole
growth slows down. At this stage, we measure the hole perimeter (indicated
by the dashed line) and its increase with further irradiation. (d)
Hole perimeter as a function of dose for various 20kV (red) and 80kV
experiments (black). Solid and dashed lines denote separate experiments
with presumably different (although uncontrolled) environmental conditions. }

\end{figure}

Several observations indicate that the hole growth is a chemical reaction
with residual gas in the vacuum. The most significant one comes from
the comparison of 20 kV and 80 kV data. Knock-on damage of graphene
edge atoms should be strongly suppressed if not stopped entirely under
20kV irradiation, compared to 80kV \cite{Kotakoski2011c}. However,
we actually measured a slightly more rapid growth of holes under the
20kV beam. Second, we found that the growth rate of holes strongly
varied in different experiments, in spite of similar irradiation conditions.
Parameters that varied between these experiments were the sample holder,
the pumping time after mounting the sample, or the use of the cold
trap. In particular, we observed that hole growth slowed down after
extended pumping (several hours after inserting the sample), compared
to immediately after inserting the sample. We also noted one case
of a leaking sample holder (contaminated o-ring) where hole growth
was too rapid for achieving any high-resolution images. Indeed, exposure
to a low-voltage electron beam under a controlled atmosphere of water
or oxygen was used previously for a controlled cutting of carbon materials
\cite{Yuzvinsky2005}. Etching of carbon with residual gas in the
TEM column was also described by other authors \cite{Molhave2007}.
Secondary displacements (e.g. displaced H atoms that, in turn, displace
C atoms) also can not play a signifcant role, as again one would expect
a clear dependence on acceleration voltage for this mechanism. In
summary, we conclude that hole formation and growth depends on the
contamination on the sample (which acts as a catalyst) and on the
residual vacuum levels, and these effects dominate over the knock-on
displacements under our experimental conditions.

\section{Description of the supplementary videos}

~\\
\emph{Supplementary video S1:} A side-by-side comparison of the
$^{12}\textrm{C}$ and $^{13}\textrm{C}$ graphene sample under 95~keV
electron irradiation, recorded under identical imaging conditions.
One can clearly see the much more rapid decay of the $^{12}\textrm{C}$
sample in the electron beam. The total dose at the end of the image
sequence is $1.7\cdot10^{9}\frac{e^{-}}{\textrm{nm}^{2}}$.

~\\
\emph{Supplementary video S2:} A long image sequence showing a
$^{12}\textrm{C}$ graphene membrane under 80~keV electron irradiation.
No change can be detected within the pristine lattice area. The total
dose at the end of the image sequence is $2\cdot10^{9}\frac{e^{-}}{\textrm{nm}^{2}}$.

~\\
\emph{Supplementary video S3:} A long image sequence showing a
$^{12}\textrm{C}$ graphene membrane under 80~keV electron irradiation
(showing a larger area view as compared to video S2). No change can
be detected within the pristine lattice area, but the formation of
extended holes at the contamination spots is clearly evident. The
total dose at the end of the image sequence is $7\cdot10^{8}\frac{e^{-}}{\textrm{nm}^{2}}$.
In particular, holes start at the heavier (non-carbon) contamination,
which appears to catalyze the reaction.

~\\
\emph{Supplementary video S4:} Formation of defects in $^{12}\textrm{C}$
graphene under 100~keV electron irradiation. The total dose at the
end of the image sequence is $1.5\cdot10^{9}\frac{e^{-}}{\textrm{nm}^{2}}$.
A few seconds (less than 30s, total time of the sequence is ca. 1
hour) are missing after frames 10, 35, 91 and 132 as the focus was
readjusted. 

~\\
\emph{Supplementary video S5:} Formation of defects in $^{13}\textrm{C}$
graphene under 100~keV electron irradiation (note that frames 120-177
are out of focus; frames 198-213 are recorded under inverse conditions
i.e. atoms appear white). The total dose at the end of the image sequence
is $6\cdot10^{9}\frac{e^{-}}{\textrm{nm}^{2}}$.

\bibliographystyle{ieeetr}
\bibliography{library,franz}

\end{document}